\documentclass[aps,pra,twocolumn,groupedaddress,amsmath,amsfonts,longbibliography]{revtex4-2}

\usepackage{graphicx}
\usepackage{braket}
\usepackage{bm}
\usepackage{hyperref}

\usepackage{xcolor}

\begin{document}

\title{Nonadiabatic effects in attosecond transient absorption spectroscopy}

\author{Nikolaj S. W. Ravn}
\affiliation{Department of Physics and Astronomy, Aarhus University, 8000 Aarhus C, Denmark}

\author{Lars Bojer Madsen}
\affiliation{Department of Physics and Astronomy, Aarhus University, 8000 Aarhus C, Denmark}

\date{\today}

\begin{abstract}
We study effects of nonadiabatic couplings in a model of a diatomic molecule in the context of attosecond transient absorption spectroscopy. By using a model system consisting of four diabatic electronic states and with a variable strength of the diabatic coupling, we can explore attosecond transient absorption spectra in different regimes of the vibronic couplings between the electronic and nuclear dynamics, and determine when nonadiabatic couplings can be ignored or when they have a substantial effect. The findings are rationalized in terms of a multilevel model, which captures aspects of both electronic and nuclear degrees of freedom.
\end{abstract}

\maketitle

\section{Introduction}

Attosecond transient absorption spectroscopy (ATAS) is an experimental technique used to study the dynamics of electrons and nuclei on their natural time scales. ATAS is a pump-probe type method, where a sub-femtosecond extreme ultraviolet (XUV) pulse together with a few-cycle infrared (IR) pulse is used. Varying the time delay between the two pulses gives access to temporal information of the induced dynamics. ATAS offers both temporal and spectral resolution as the delay between the two pulses can be controlled to a high degree. The downside of the method being that time enters indirectly through the time delay, as the recorded spectrum is a time integrated signal. For recent reviews on ATAS see Refs.~\cite{Borja:16,Wu_2016,doi:10.1098/rsta.2017.0463}.

In atoms, ATAS  has been used to observe the motion of valence electrons in Kr~\cite{goulielmakis2010}, autoionization in Ar~\cite{wang2010attosecond}, AC Stark shift in Kr~\cite{Wirth195} and He~\cite{PhysRevLett.109.073601}, electron wave packets in He~\cite{PhysRevLett.106.123601,ott2014} and dynamics in Xe~\cite{Sabbar2017,PhysRevA.95.031401}. Typical features of ATA spectra~\cite{PhysRevA.96.013430} include Autler-Towes splitting of absorption lines~\cite{PhysRevA.85.053422,PhysRevA.88.043416}, oscillating fringes around bright states coming from three-photon interference~\cite{PhysRevA.87.033408,Chini_2014}, light-induced structures (LIS) around dark states attributed to two-photon interference~\cite{PhysRevA.86.063408} and hyperbolic sidebands around the main absorption lines associated with free-induction decay (see, e.g., the discussion in Ref.~\cite{PhysRevA.96.013430}).

In ATAS of molecular systems, compared to atoms, one has to take into account the nuclear vibrational  dynamics~\cite{PhysRevA.91.043408,PhysRevA.92.023407} as well as alignment~\cite{PhysRevA.94.043414} in calculations of the spectra. Theory and experiment have mostly focused on diatomic systems such as H$_2$~\cite{PhysRevA.94.023403}, N$_2$~\cite{doi:10.1021/acs.jpca.5b11570,warrick2017} and O$_2$~\cite{PhysRevA.95.043427}. Work has also been done on systems with a permanent dipole~\cite{PhysRevA.98.053401,Drescher_2020}, which affects the nuclear dynamics since vibrational transitions within a given electronic state are now possible. So far effects of nonadiabatic transitions have not been discussed in detail for diatomic molecules, i.e, so far the vibronic couplings between adiabatic Born-Oppenheimer curves have been neglected and coupling between curves only occurs due to the coupling with the external laser pulses.

The description of the molecular systems is usually done by expanding the full wave packet of the system over a known set of electronic states, which translates the problem into a description of the resulting nuclear wave packets moving on the electronic potential-energy curves. In an ab initio treatment one usually starts in the adiabatic representation with the adiabatic molecular properties obtained through a quantum chemistry calculation. If the Born-Oppenheimer approximation can be assumed valid, the nonadiabatic effects can be ignored and only the potential-energy curves and dipole moments are needed for a simulation of the nuclear dynamics. However, avoided crossings, around which the nonadiabatic effect can seldom be neglected, are abundant in many molecules and can alter the nuclear dynamics significantly. It is the purpose of the present work to analyze aspects of the effects of nonadiabatic couplings on ATAS of diatomic molecules.

The paper is organized as follows. In Sec.~\ref{Sec:Theory}, we summarize the theory for ATAS in diatomic molecules and how this is implemented numerically. In Sec.~\ref{Sec:AdiabDiab}, we go into details with the adiabatic and diabatic representations, with a focus on the nonadiabatic effects and the avoided crossings coming from the vibronic coupling between the vibrational and electronic dynamics. In Sec.~\ref{Sec:Model}, we introduce our model system, which is defined in the diabatic representation where the curves are allowed to cross and the vibronic coupling is defined by the diabatic coupling element, and describe how we calculate the corresponding adiabatic curves. In Sec.~\ref{Sec:Results}, we first consider the simpler case of XUV-only spectroscopy, which we use to characterize the curves based on the strength of the diabatic coupling and to discuss the induced nuclear motion. In Sec.~\ref{Sec:Results}, we further analyse the full ATA spectra of the model in three distinct regimes defined by the diabatic coupling strength. In Sec.~\ref{Sec:Analytical}, we extend the analytical three-level model of Ref.~\cite{PhysRevA.96.013430} to a multilevel model capturing aspects of nuclear dynamics to discuss the description of the different features in the ATA spectrum. Finally, Sec.~\ref{Sec:Conclusion} concludes the work. Atomic units, [a.u.], are used throughout, unless stated otherwise.

\section{Theory for ATAS}
	\label{Sec:Theory}

In our convention the IR pulse is held fixed in time with its center at $t=0$ and with the center of the XUV pulse given by the time delay $\tau$. Hence for positive delay, the XUV pulse arrives after the center of the IR pulse, while for negative delay the XUV arrives before the center of the IR. In our description we assume a single-system approximation, where the macroscopic propagation effects are only accounted for in an approximative manner. It has been argued that this is a valid approximation for dilute gases~\cite{PhysRevA.83.013419}. In this case, the ATA spectrum can be described by the response function~\cite{PhysRevA.91.043408}
	\begin{equation}
		S(\omega,\tau) = \frac{4\pi\rho\omega}{c}
			\textrm{Im} \big[ \tilde{F}_{\text{XUV}}^*(\omega,\tau) \tilde{D}(\omega,\tau) \big],
		\label{eq:ResponseFunc}
	\end{equation}
with $\tilde{F}_{\text{XUV}}(\omega,\tau)$ the Fourier transform of the time-dependent incoming attosecond XUV field, $\tilde{D}(\omega,\tau)$ the Fourier transform of the time-dependent induced dipole moment of the system, and $\rho$ the density of the target gas, which is arbitrarily set to $\rho=1$. In our work, we use the following definition for the Fourier transform $\tilde{f}(\omega) = \frac{1}{\sqrt{2\pi}} \int_{\mathbb{R}} dt \, f(t) e^{-i\omega t}$. The derivation of Eq.~\eqref{eq:ResponseFunc} can be found in Refs.~\cite{PhysRevA.91.043408, PhysRevA.85.013415}, see also Ref.~\cite{Baggesen_2011} for a discussion in the context of high-order harmonic generation. In Ref.~\cite{PhysRevA.91.043408} the relation between different expressions for the single-system response function is discussed. The conclusion from that discussion is that $S(\omega,\tau)$ of Eq.~\eqref{eq:ResponseFunc} is a good measure of the modification of the XUV spectrum when the generated field is weak. In our convention a negative value of $S(\omega,\tau)$ indicates absorption of light, while a positive value indicates emission.

The incoming fields are defined through the relation $F(t) = - \partial_t A(t)$, with the vector potential
	\begin{equation}
		A(t) = A_0 \exp \bigg[ - \frac{(t-t_c)^2}{T^2/4} \bigg]
			\cos \big[ \omega(t-t_c) \big].
		\label{eq:VectorPotential}
	\end{equation}
Both fields are linearly polarized; with polarization axis parallel to the internuclear axis of the diatomic system. The construction of the electric field from the vector potential as specified above ensures that the short pulses include no unphysical DC components~\cite{PhysRevA.65.053417}. The parameters for the XUV and IR pulses to be used in the calculations are collected in Table~\ref{Table:FieldParams}. The intensity in atomic units is obtained by dividing the intensity-values of Table~\ref{Table:FieldParams} by $3.5\times10^{16}$~W/cm$^2$. In atomic units the magnitude of the vector potential in Eq.~\eqref{eq:VectorPotential} is obtained as $A_0 = \sqrt{I}/\omega$.

	\begin{table}
	\caption{Parameters of XUV and IR pulses defined by the vector potential in Eq.~\eqref{eq:VectorPotential}, and used for the calculations in the present work.}
	\label{Table:FieldParams}
	\begin{ruledtabular}
	\begin{tabular}{lcc}
											&	\textrm{XUV}	& \textrm{IR}	\\ \colrule
		Central wave length $\lambda$ [nm]	&	84				&	3200		\\
		Central frequency $\omega$ [eV]		&	14.7			&	0.386		\\
		Intensity $I$  [W/cm$^2$]			&	$5\times10^{7}$	&	$10^{13}$	\\
		Cycles $N_c$						&	2				&	3			\\
		Duration $T = N_c \frac{2\pi}{\omega}$ [fs]
											&	0.56			&	32		
	\end{tabular}
	\end{ruledtabular}
	\end{table}

To describe the diatomic molecular system, we employ an $N$-surface model, where the full wave packet of the system is expanded in $N$ electronic states $\Phi_n(\mathbf{r};R)$ (either adiabatic or diabatic, to be discussed and specified in Sec.~\ref{Sec:AdiabDiab})
	\begin{equation}
		\Psi(R,\mathbf{r},t)
			= \sum_n^N \chi_n(R,t) \Phi_n(\mathbf{r};R),
		\label{eq:NucExpansion}		
	\end{equation}
with $\mathbf{r}$ denoting the set of all electron coordinates. The corresponding nuclear wave packets $\chi_n(R,t)$ then satisfy the following set of coupled differential equations
	\begin{equation}
	 i \partial_t \chi_n (t,R)
	= \sum_{k}^N \Big( H_{nk}(R) + V^{\text{L}}_{nk}(t,R) \Big) \chi_k (t,R),
		\label{eq:NucEq}
	\end{equation}
with $H_{nk}(R)$ being the field-free Hamiltonian of the system containing the kinetic and potential-energy terms for the vibrational motion. The rotational degree of motion of the molecule is ignored due to its much longer time scale and the nuclear degrees of freedom are then solely described by the internuclear distance $R$. In Eq.~\eqref{eq:NucEq}, $V^{\text{L}}_{nk}(t,R)$ denotes the interaction with the incoming pulses. In the length gauge and assuming the dipole approximation we have
	\begin{equation}
		V^{\text{L}}_{nk}(t,R) = -F(t) D_{nk}(R),
	\end{equation}
with $F(t) = F_{\text{XUV}}(t) + F_{\text{IR}}(t)$ the total incoming field and $D_{nk} = \braket{\Phi_n \vert \hat{D} \vert \Phi_k}_{\mathbf{r}} = \int \! d\mathbf{r} \, \Phi_n^*(\mathbf{r};R) \hat{D} \Phi_k (\mathbf{r};R)$ being the dipole moments of the electronic states. Here we let $\braket{\cdot\vert\cdot}_{\mathbf{r}}$ denote integration over the electronic coordinates. We will adopt the notation of electronic states being bright with respect to one another if the dipole moment $D_{nk}$ between them is non-zero, and being dark if the dipole moment between them is zero. The total time-dependent dipole moment of the molecule used to calculate the response function in Eq.~\eqref{eq:ResponseFunc} can then be written
	\begin{equation}
		\braket{D} \! (t) = \sum_{n,k}^N \int \! dR \, \chi_n^*(R,t) D_{nk}(R) \chi_k(R,t).
		\label{eq:TimeDepDipole}
	\end{equation}

The coupled nuclear equations in Eq.~\eqref{eq:NucEq} are propagated using a split-step Fast-Fourier-transform method with fixed time step $\Delta t = 0.05$~a.u. $=1.21$~as, box size $R_{\text{max}} = 10$~a.u. and grid spacing $\Delta R = R_{\text{max}} /N_R$ with the number of grid points $N_R = 2^{10}$. The boundary of the box is equipped with a complex adsorbing potential (CAP), which is added to the diagonal part of the potentials $V(R) \rightarrow V(R) + V_{\text{CAP}}(R)$. The CAP is used to remove unphysical reflection of outgoing wave packets by absorbing them at the boundary. In the present work we use a potential of the form
	\begin{equation}
		V_{\text{CAP}}(R) =
			\begin{cases}
				-i\eta (R - R_{\text{CAP}} )^2, 	& \quad \text{for } R_{\text{CAP}} < R \\
				0,								& \quad \text{otherwise}, 
			\end{cases}
	\end{equation}
with amplitude $\eta = 10^{-3}$ and the beginning of the potential at $R_{\text{CAP}} = 8$~a.u. The initial ground state is found by field-free imaginary time propagation. Results have been checked for convergence by independently decreasing time step, $\Delta t$, by 20~\%, increasing box size, $R_{\text{max}}$, by 20~\% and decreasing grid spacing, $\Delta R$, by 20~\%. These changes in calculational parameters led to no visible changes in the results to be presented below.

Before Fourier transformation, the time-dependent dipole moment in Eq.~\eqref{eq:TimeDepDipole} is multiplied by a window function
		\begin{equation}
			W(t-t_0) = 
			\begin{cases}
				1,
					& t < t_0 \\
				\exp \Big[ - \frac{(t-t_0)^2}{T_0^2/4} \Big],
					& t_0 \leq t,
			\end{cases}
			\label{eq:WindoWFunc}
		\end{equation}
which is used to mimic the dephasing of the dipole moment seen in a real experimental setting. In our case the start of the window function coincide with the center of the XUV $t_0=\tau$ and the width $T_0 = 290.8$~fs.

\section{Adiabatic and diabatic representations}
	\label{Sec:AdiabDiab}

In this section, to make the presentation self-contained, we present a brief summary of results regarding equations of motion for the nuclear wave packets in the adiabatic and diabatic representations.
In Eq.~\eqref{eq:NucExpansion} the full wave packet of the system is expanded over a basis of electronic states, which we assume is known. A common choice of basis is the adiabatic electronic states $\Phi_n^{(a)}$ defined to satisfy the time-independent Schr{\"o}dinger equation for fixed adiabatic parameter $R$
	\begin{equation}
		H_{\text{e}}(R) \Phi_n^{(a)}(\mathbf{r};R)
			= V_n^{(a)}(R) \Phi_n^{(a)} (\mathbf{r};R).
		\label{eq:AdiabEigen}
	\end{equation}
Here the electronic Hamiltonian is $H_{\text{e}}(R) = T_{\text{e}} + V_{\text{NN}}(R) + V_{\text{Ne}}(R,\mathbf{r}) + V_{\text{ee}}(\mathbf{r})$, consisting of the electronic kinetic energy and the Coulomb interaction between the two nuclei, between the nuclei and the electrons and between the electrons.  In this representation the corresponding Hamiltonian governing the nuclear dynamics in Eq.~\eqref{eq:NucEq} reads
	\begin{align}
		H^{(a)}_{nk}(R) 
			&= \Big( - \tfrac{1}{2 \mu_{\text{N}}} \tfrac{\partial^2}{\partial R^2} + V^{(a)}_n(R) \Big) \delta_{nk} \nonumber\\
			& \quad - \tfrac{1}{2\mu_{\text{N}}} \Big( 2 P_{nk}(R) \tfrac{\partial}{\partial R} + Q_{nk}(R) \Big).
		\label{eq:ExactAdiabHam}
	\end{align}
The representation in Eq.~\eqref{eq:ExactAdiabHam} is often employed since the adiabatic potentials $V_n^{(a)}$ are diagonal and are naturally obtained from quantum chemistry calculations considering the electronic part of the problem for fixed $R$, Eq.~\eqref{eq:ExactAdiabHam}. The downside with this representation is the presence of the two nonadiabatic  couplings
	\begin{equation}
		P_{nk}(R) = \big\langle \Phi_n^{(a)} \big\vert \tfrac{\partial}{\partial R} \Phi_k^{(a)} \big\rangle_{\mathbf{r}}
	\end{equation}
and
	\begin{equation}
		Q_{nk}(R) = \big\langle \Phi_n^{(a)} \big\vert\tfrac{\partial^2}{\partial R^2} \Phi_k^{(a)} \big\rangle_{\mathbf{r}},
	\end{equation}
which arise from the dependence of the adiabatic electronic states in Eq.~\eqref{eq:AdiabEigen} on the internuclear distance $R$. These nonadiabatic terms represent the vibronic couplings (VCs), which describe the interaction between the nuclear and electronic degrees of freedom. Equations~\eqref{eq:NucEq} and~\eqref{eq:ExactAdiabHam} show that without nonadiabatic couplings, there are no transitions between the adiabatic potential-energy curves without the presence of an external field. In the adiabatic Born-Oppenheimer approximation, the nonadiabatic terms are neglected, and the Hamiltonian in Eq.~\eqref{eq:ExactAdiabHam} reduces to only the nuclear kinetic energy and the diagonal adiabatic potential-energy curves. This is often a good approximation, but it breaks down near avoided crossings, where the nonadiabatic couplings can be substantial, see, e.g., Refs.~\cite{PhysRev.179.111,doi:10.1063/1.440893}. Physically, this occurs because close to the avoided crossings the energy difference between the electronic adiabatic potential-energy curves decreases and becomes comparable to the energy differences associated with the nuclear degree of freedom. Hence the timescales for electronic and nuclear motion become comparable, the foundation for the Born-Oppenheimer approximation breaks down, and the nuclear and electronic degrees of freedom are mixed strongly via the vibronic couplings. This consideration raises a question on the validity of neglecting the nuclear rotational degrees of freedom. The rotational dynamics are usually disregarded as their associated time scales are much longer, and the corresponding energy splitting much smaller, than those of the electronic and vibrational dynamics. However, for avoided crossings with very narrow gaps in energy, the electronic splitting in energy can become comparable to the splitting of the rotational energy levels, and a coupling between the respective degrees of freedom can be significant. In the present work, we consider a scan over a wide range of different vibronic coupling strengths, and may hence include scenarios where ignoring the rotational degrees of freedom cannot be strictly justified. However, the main interest in this study is in the description of intermediate vibronic coupling, and the extreme cases of strong and weak regimes, where the problem may occur, are used for comparisons. We will, therefore, ignore the influence of the rotational dynamics, but keep in mind that further work on its influence can be of interest. 

The nonadiabatic couplings describe correlated electron-nuclear dynamics, and in this sense account for a more richer dynamics. The form of $P_{nk}$ and $Q_{nk}$, however, can make a numerical treatment of the equations of motion following Eqs.~\eqref{eq:ExactAdiabHam} and~\eqref{eq:NucEq} complicated. The two nonadiabatic terms can be removed by the use of a unitary transform of the nuclear wave packets into the diabatic representation
	\begin{equation}
		\chi_n^{(a)}(R,t)
			= \sum_k^N U_{nk}(R) \chi_k^{(d)}(R,t), 
	\end{equation}
where the transformation matrix is found by solving the equation~\cite{PhysRev.179.111}
	\begin{equation}
		\tfrac{\partial}{\partial R} \mathbf{U}
			= - \mathbf{P} \mathbf{U}.
			\label{eq:DiffEqTransform}
	\end{equation}
The corresponding potential energy and dipole moments in the diabatic representation are found by
	\begin{equation}
		\mathbf{U}^\dagger \mathbf{V}^{(a)} \mathbf{U}
			= \mathbf{V}^{(d)}
			\quad \text{and} \quad
		\mathbf{U}^\dagger \mathbf{D}^{(a)} \mathbf{U}
		= \mathbf{D}^{(d)}.
		\label{Eq:TransfromPotDip}
	\end{equation}
The price of removing the nonadiabatic couplings is that the potential energies are no longer diagonal, and the nuclear Hamiltonian becomes
	\begin{equation}
		H^{(d)}_{nk}(R)
			= -\tfrac{1}{2\mu_{\text{N}}} \tfrac{\partial^2}{\partial R^2} \delta_{nk} + V^{(d)}_{nk}(R).
		\label{eq:ExactDiabHam}
	\end{equation}
In the diabatic representation the avoided crossings are removed and the couplings between the curves are captured in the off-diagonal diabatic potential terms and the altered shape of the diagonal potentials, which are now crossing. The transformation from adiabatic to diabatic defined by Eq.~\eqref{eq:DiffEqTransform} is only exact in the case of a complete basis of electronic states. For a truncated basis, such as the case we consider in the present work, the adiabatic and diabatic representations are only approximately equivalent. This is similar to the freedom of choice for the form of the interaction with the electromagnetic field, where, e.g., the velocity and length gauges are only exactly equivalent when a complete basis is considered.

\section{Model system}
	\label{Sec:Model}

\begin{figure}
		\centering
		\includegraphics[width=0.25\textwidth]{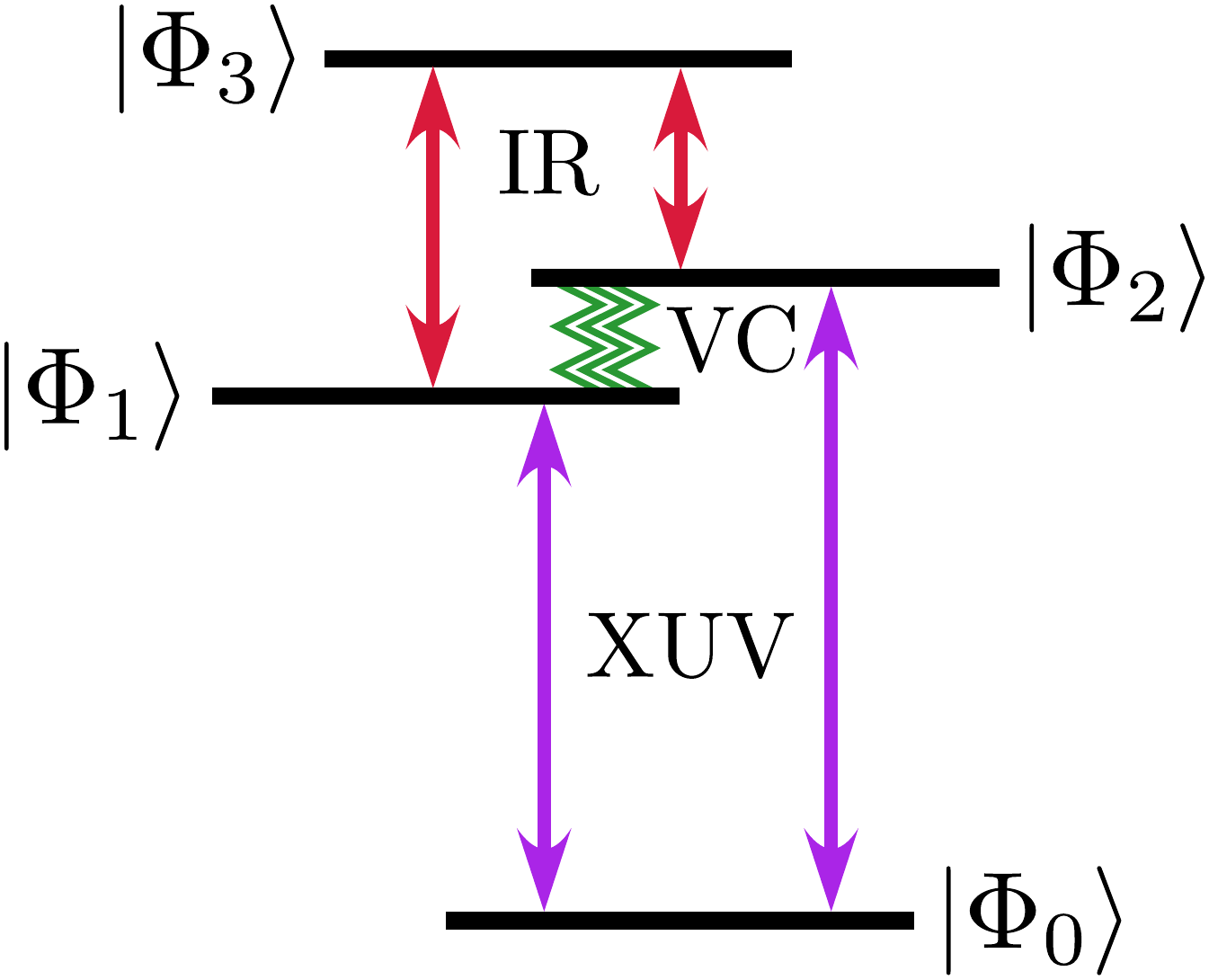}
		\caption{Level scheme illustrating the coupling between the different electronic states of the model system. The ground state $\ket{\Phi_0}$ has a dipole-allowed transition to the two lowest excited states $\ket{\Phi_1}$ and $\ket{\Phi_2}$ and these two have an allowed coupling to the third excited state $\ket{\Phi_3}$. Due to the difference in energy and the frequency and intensity of the fields, the coupling from the ground state is mainly by the XUV pulse while the coupling between the excited states is mediated by the IR pulse. We further have a vibronic coupling (VC), describing the nonadiabatic coupling between the two lowest excited states. The electronic states can represent both diabatic or adiabatic states, see discussion in Sec.~\ref{Sec:Model}.}
		\label{Fig:LevelScheme}
	\end{figure}

In the present work we consider a model system consisting of four electronic states, a ground state $\ket{\Phi_0}$ and three excited states $\ket{\Phi_1}$, $\ket{\Phi_2}$ and $\ket{\Phi_3}$. The ground state has a non-zero dipole coupling to the two excited states $\ket{\Phi_1}$ and $\ket{\Phi_2}$  and these two states further have a coupling to the excited state $\ket{\Phi_3}$, see Fig.~\ref{Fig:LevelScheme}. In addition, we introduce a vibronic coupling between the two excited states $\ket{\Phi_1}$ and $\ket{\Phi_2}$. This level scheme captures essential characteristics of the case of a homonuclear diatomic molecule, where vibronic couplings can only occur between states of the same symmetry and dipole couplings only between states of different symmetry. The present model can therefore represent a system where $\ket{\Phi_0}$ and $\ket{\Phi_3}$ belong to one symmetry and $\ket{\Phi_1}$ and $\ket{\Phi_2}$ belong to another.

\begin{figure}
		\centering
		\includegraphics[width=0.45\textwidth]{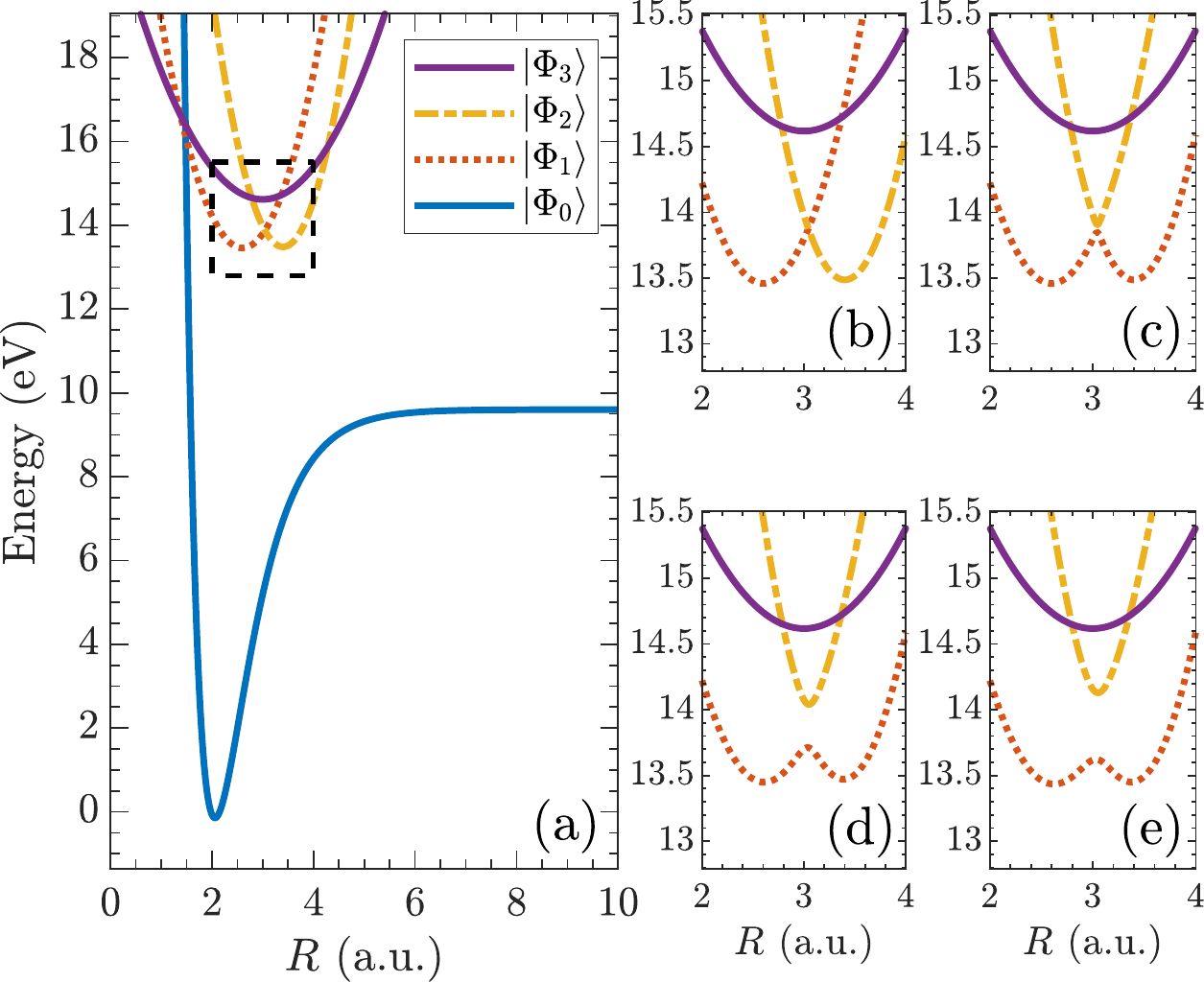}
		\caption{Potential energy curves of the 4 electronic states in the model system. (a) Diagonal diabatic potentials defined in Eqs.~\eqref{Eq:MorseGround} and~\eqref{Eq:HarmExcited}. (b) Zoomed-in view of the excited diabatic potentials around the crossing. (c) Adiabatic curves for weak diabatic coupling ($V_{12,0}^{(d)}=0.0275$~eV). (d) Adiabatic curves for intermediate diabatic coupling ($V_{12,0}^{(d)}=0.1649$~eV). (e) Adiabatic curves for strong diabatic coupling ($V_{12,0}^{(d)}=0.2542$~eV). In the plots the potentials are shifted in energy such that the vibrational ground state of the ground state curve is at zero energy, facilitating a direct comparison with the energy scales in the spectrograms.}
		\label{Fig:Potentials}
	\end{figure}

We define the potential curves and the dipole moments initially in the diabatic representation, as the potential-energy curves are allowed to cross and the dipole moments are well behaved over the crossing~\cite{PhysRev.179.111}. Some parameters of the model are chosen such that they resemble the N$_2$ molecule, i.e., we use the reduced nuclear mass of N$_2$ and choose curves with harmonic frequencies comparable to those in N$_2$.  For the ground state curve we use a Morse potential
	\begin{equation}
		V_{00}^{(d)}(R) = D_e \Big( 1 - e^{-a(R-R_0)} \Big)^2,
		\label{Eq:MorseGround}
	\end{equation}
with parameters $D_e$, $a$, $R_0$, such that it agrees with the experimental available harmonic and anharmonic frequencies and ionization energy for N$_2$~\cite{cccbdb}.The three excited states are chosen as energy displayed parabolas
	\begin{equation}
		V_{nn}^{(d)}(R) = \tfrac{1}{2} \mu_{\text{N}} \omega_n^2 \big(R - R_{0,n} \big)^2 + b_n.
		\label{Eq:HarmExcited}
	\end{equation}
with the reduced nuclear mass $\mu_{\text{N}}$, frequencies $\omega_1 = 0.095$~eV, $\omega_2 = 1.2\,\omega_1$, $\omega_3 = 0.6\,\omega_1$, energy offset $b_1 = 13.6$~eV, $b_2 = b_1 + 0.027$~eV, $b_3 = b_1 + 3\,\omega_{\text{IR}}$ and minima $R_{0,1} = 2.6$~a.u., $R_{0,2} = 3.4$~a.u. and $R_{0,3} = R_c = 3.0$~a.u. respectively. These latter parameters were chosen based on some typical values from excited curves, see Ref.~\cite{warrick2017}. We further have an off-diagonal diabatic coupling between the two lowest-excited states centered at the crossing $R_c$
	\begin{equation}
		V_{12}^{(d)}(R) = V_{12,0}^{(d)} e^{-(R-R_c)^2},
		\label{eq:DiabaticCoupling}
	\end{equation}
where we use a Gaussian with a fixed width but variable amplitude $V_{12,0}^{(d)}$. The Gaussian is similar to the one used in Ref.~\cite{PhysRevLett.117.043201}. Comparisons of the corresponding adiabatic curves and nonadiabatic couplings show little dependence on changes in the width, and we therefore keep it fixed.  In Ref.~\cite{BeakhojConical} nonadiabatic effects in ATAS at a conical intersection were considered using a model where the diabatic couplings were modelled using a low-order Taylor expansion, see Ref.~\cite{ConicalIntersections}. This latter approach has the advantage of being simple to fit to experimental parameters, but is not applicable to describe dynamics with larger amplitude of the nuclear motion, i.e., it is accurate only close to the crossing. The form in Eq.~\eqref{eq:DiabaticCoupling} puts no restriction on the motion of the nuclear wave packets. The diagonal diabatic curves are shown in Fig.~\ref{Fig:Potentials}~(a) together with a zoomed-in view of the crossing between the two excited curves in Fig.~\ref{Fig:Potentials}~(b). For the diabatic dipole moments we assume constant values of $D_{01}^{(d)} = D_{13}^{(d)} = 1.0$ a.u. and $D_{02}^{(d)} = D_{23}^{(d)} = 0.5$ a.u. Due to symmetry considerations the transformation between the adiabatic and diabatic electronic states is such that the dipole coupling scheme between the electronic states is preserved, i.e., values of the dipole moments as function of $R$ are different in the two representation, but non-zero dipole moments stay non-zero and dipole moments that are zero stay zero.

For a given value of the amplitude of the diabatic coupling $V_{12,0}^{(d)}$, the adiabatic potential matrix $\mathbf{V}^{(a)}$ can be found by diagonalization of the diabatic potential matrix $\mathbf{V}^{(d)}$. The unitary transformation matrix $\mathbf{U}$ can then be constructed from the corresponding eigenvectors, which can further be used to find the adiabatic dipole moments $\mathbf{D}^{(a)}$, see Eq.~\eqref{Eq:TransfromPotDip}. The first-order nonadiabatic coupling $\mathbf{P}$ can be found by using Eq.~\eqref{eq:DiffEqTransform}, taking the derivative of the transformation matrix using, e.g., finite difference. 

Depending on the strength of the diabatic coupling $V_{12,0}^{(d)}$, we can divide the curves into three regimes; weak, intermediate and strong diabatic coupling (see further discussion in Sec.~\ref{Sec:XUVscan}). Figure~\ref{Fig:Potentials} shows the adiabatic curves in (c) the weak regime, (d) the intermediate regime and (e) the strong regime. Performing a calculation with the adiabatic curves and including the nonadiabatic couplings would yield the same result as that obtained by working in the diabatic representation. It is of interest here to neglect the nonadiabatic couplings in the adiabatic representation to see when this is a valid approximation. In the rest of the paper we will refer to the exact set of curves and couplings in the diabatic representation as the exact diabatic curves and the set of corresponding adiabatic curves without the nonadiabatic couplings as the approximate adiabatic curves. The former represent the system exactly and the latter represent the system under the Born-Oppenheimer approximation.
By taking the derivative with respect to $R$ of the adiabatic eigenvalue equation in Eq.~\eqref{eq:AdiabEigen} and projecting onto another adiabatic electronic state, we obtain, after some rearrangement the relation
	\begin{equation}
		P_{nk}(R) = \frac{\Big\langle \Phi^{(a)}_n \Big\vert \frac{\partial H_{\text{e}}}{\partial R} \Big\vert \Phi^{(a)}_k \Big\rangle_{\mathbf{r}} }{V^{(a)}_n(R) - V^{(a)}_k(R) }.
		\label{eq:NonAdiabP}
	\end{equation}
By considering a simple two-level model of the two adiabatic potential curves near the avoided crossing, one can further relate the adiabatic splitting in energy at the crossing with the strength of the diabatic coupling $V_n^{(a)}(R_c) - V_k^{(a)}(R_c) = 2 \Big\vert V_{nk}^{(d)}(R_c) \Big\vert$, see Ref.~\cite{bransden1992charge}. Inserting this relation in Eq.~\eqref{eq:NonAdiabP} gives
	\begin{equation}
		P_{nk}(R_c) = \frac{\Big\langle \Phi^{(a)}_n \Big\vert \frac{\partial H_{\text{e}}}{\partial R} \Big\vert \Phi^{(a)}_k \Big\rangle_{\mathbf{r}} }{2 \Big\vert V^{(d)}_{nk}(R_c) \Big\vert},
	\end{equation}
which shows that at the crossing, the nonadiabatic coupling $P_{nk}$ is inversely proportional to the diabatic coupling $V_{nk}^{(d)}$. This relation can be seen directly in Fig.~\ref{Fig:Potentials}, where a weak diabatic coupling in Fig.~\ref{Fig:Potentials}~(c) gives a very narrow avoided crossing in the adiabatic curves, and the energy gap further grows as the diabatic coupling is increased in Figs.~\ref{Fig:Potentials}~(d) and~(e).

In the case of a weak nonadiabatic coupling, i.e., a strong diabatic coupling, it is therefore preferable to work in the adiabatic representation, where the Born-Oppenheimer approximation is valid, and one can neglect the nonadiabatic couplings. In the case of a strong nonadiabatic coupling on the other hand, the diabatic coupling is weak and working in the diabatic representation and neglecting the diabatic coupling is preferred. For adiabatic states with a strong nonadiabatic coupling one can therefore perform a ``crude'' diabatization, where the adiabatic curves are connected through the crossing and one obtains an approximation of the diabatic curves without accounting for the weak diabatic coupling.

\section{Results and discussion}
	\label{Sec:Results}

\subsection{XUV-only scan}
\label{Sec:XUVscan}

	\begin{figure}
		\centering
		\includegraphics[width=0.4\textwidth]{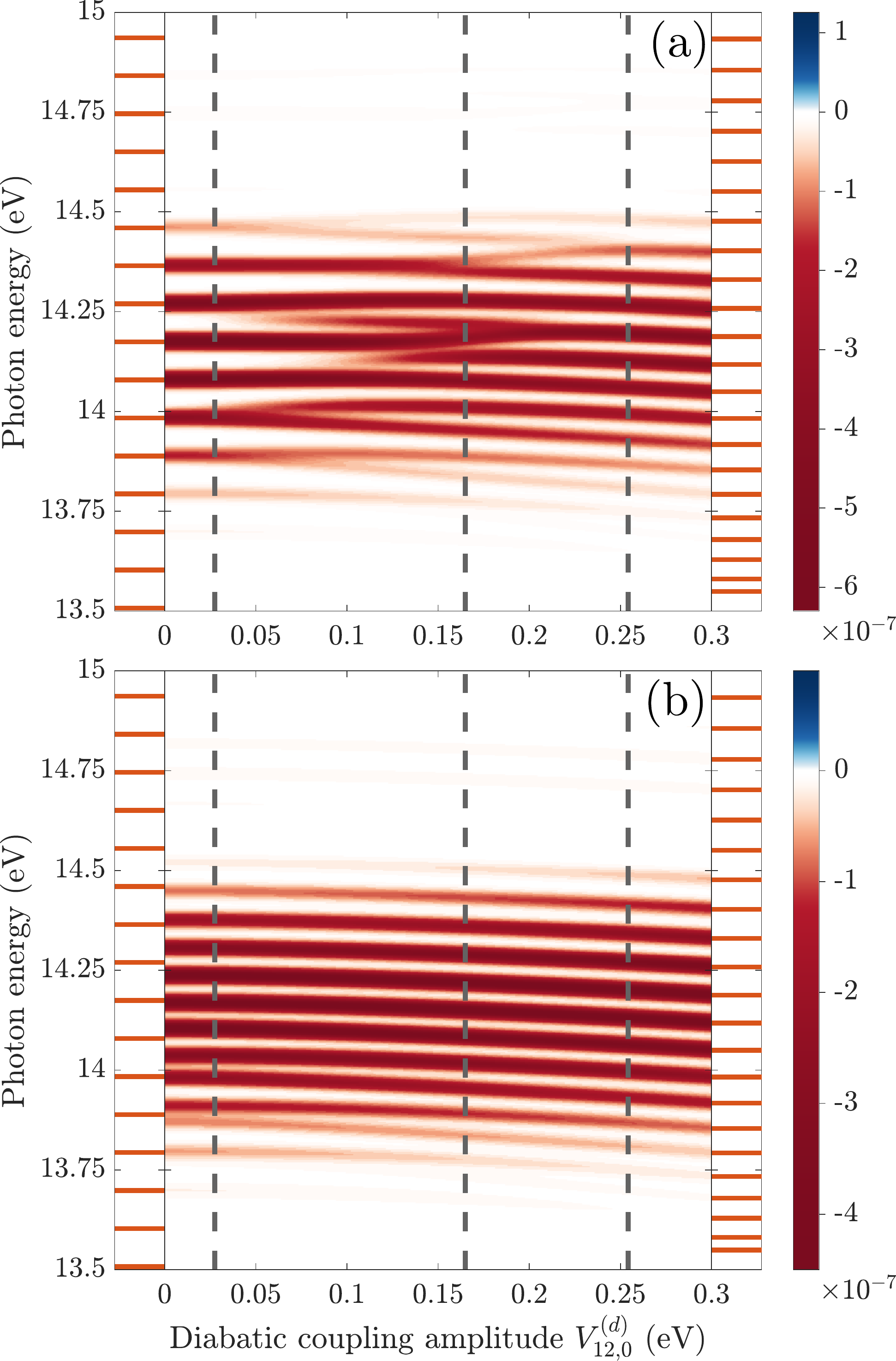}		
		\caption{Single-color XUV-only absorption spectra calculated using Eq.~\eqref{eq:ResponseFunc} as a function of the diabatic coupling amplitude $V_{12,0}^{(d)}$, see Eq.~\eqref{eq:DiabaticCoupling}. (a) Spectrum following the use of the exact diabatic curves and couplings, (b) Spectrum following the use of the adiabatic curves without the nonadiabatic couplings. The three vertical dashed lines indicate specific values of the diabatic coupling in the three regimes of weak ($V_{12,0}^{(d)} = 0.0275$~eV), intermediate ($V_{12,0}^{(d)} = 0.1649$~eV) and strong coupling ($V_{12,0}^{(d)} = 0.2542$~eV). The vibrational levels of the diabatic state $\ket{\Phi_1^{(d)}}$ for $V_{12,0}^{(d)} = 0$~eV is shown to the left of the figure and the vibrational levels of the adiabatic state $\ket{\Phi_1^{(a)}}$, with no nonadiabatic coupling, for $V_{12,0}^{(d)}= 0.3$~eV is shown to the right.}
		\label{Fig:XUVonlyScan}
	\end{figure}

	\begin{figure}
		\centering
		\includegraphics[width=0.45\textwidth]{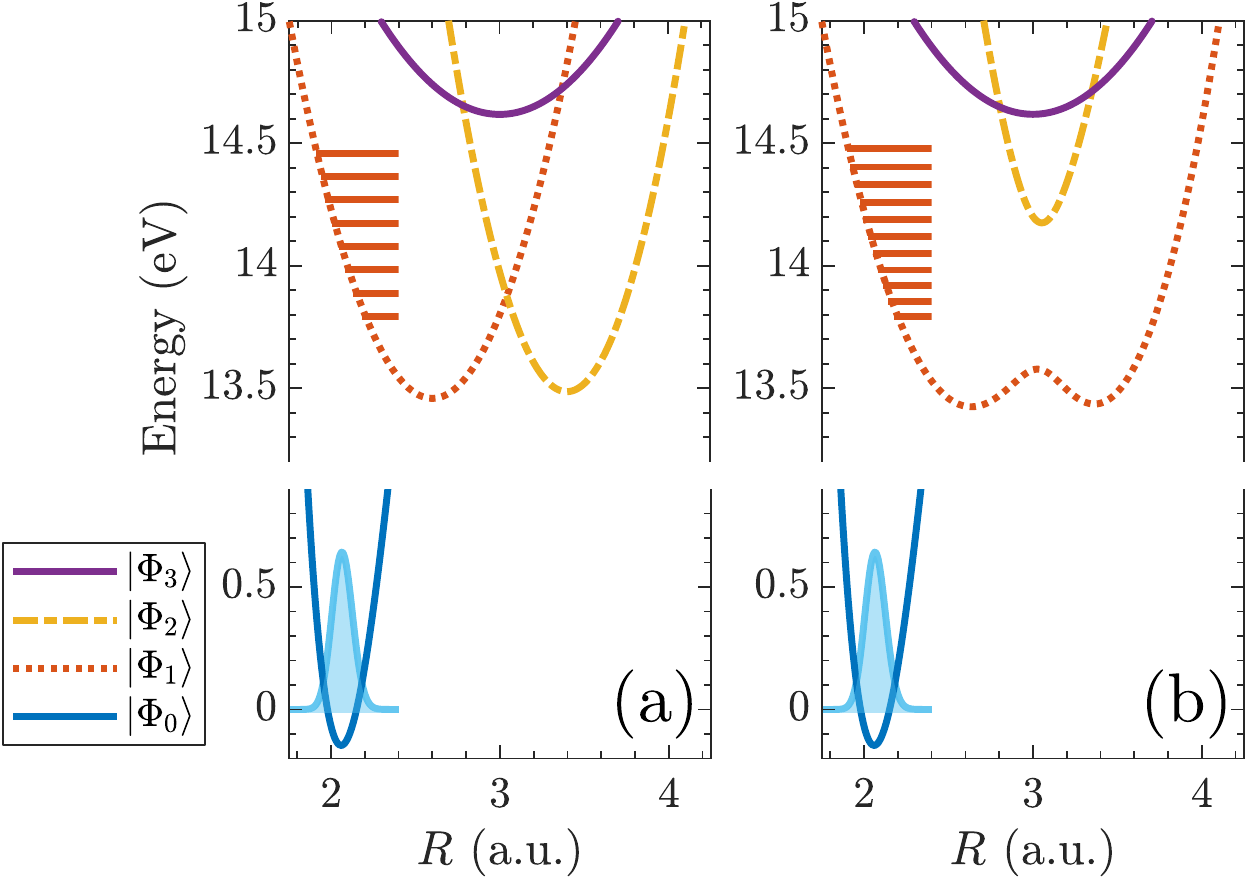}
		\caption{(a) Diabatic potentials for zero diabatic coupling $V_{12,0}^{(d)} = 0$~eV together with the vibrational levels of the diabatic state $\ket{\Phi_1^{(d)}}$, corresponding to the absorption lines in Fig.~\ref{Fig:XUVonlyScan}. (b) Adiabatic potentials for diabatic coupling $V_{12,0}^{(d)} = 0.3$~eV, together with the vibrational levels of the adiabatic state $\ket{\Phi_1^{(a)}}$, corresponding to the absorption lines in Fig.~\ref{Fig:XUVonlyScan}.}
		\label{Fig:PotentialsVibLevels}
	\end{figure}

We first consider the simplified case of an XUV-only spectrogram. In this case there is no explicit time dependence of the signal, and we can instead perform a scan over the amplitude $V_{12,0}^{(d)}$ of the diabatic coupling, see Eq.~\eqref{eq:DiabaticCoupling}. We will first consider these spectrograms in Fig.~\ref{Fig:XUVonlyScan} and later compare these to the nuclear dynamics obtained for specific values of the diabatic coupling, $V_{12,0}^{(d)}$.

In Fig.~\ref{Fig:XUVonlyScan}~(a) we show the XUV-only scan of the exact diabatic curves and couplings. Based on the resulting spectrum, we can divide the strength of the diabatic coupling into three different regimes. The three vertical lines in Fig.~\ref{Fig:XUVonlyScan}~(a) indicate specific values of $V_{12,0}^{(d)}$ in each of the three cases. We further show the vibrational levels of the first excited diabatic state, $\ket{\Phi_1^{(d)}}$, for $V_{12,0}^{(d)} = 0$~eV in the left panel and the vibrational levels of the first excited adiabatic state, $\ket{\Phi_1^{(a)}}$, for $V_{12,0}^{(d)} = 0.3$~eV and no nonadiabatic coupling in the right panel. Figure~\ref{Fig:PotentialsVibLevels} highlights the differences in the curves and vibrational level structure for these two latter cases.

	\begin{figure*}
		\centering
		\includegraphics[width=0.95\textwidth]{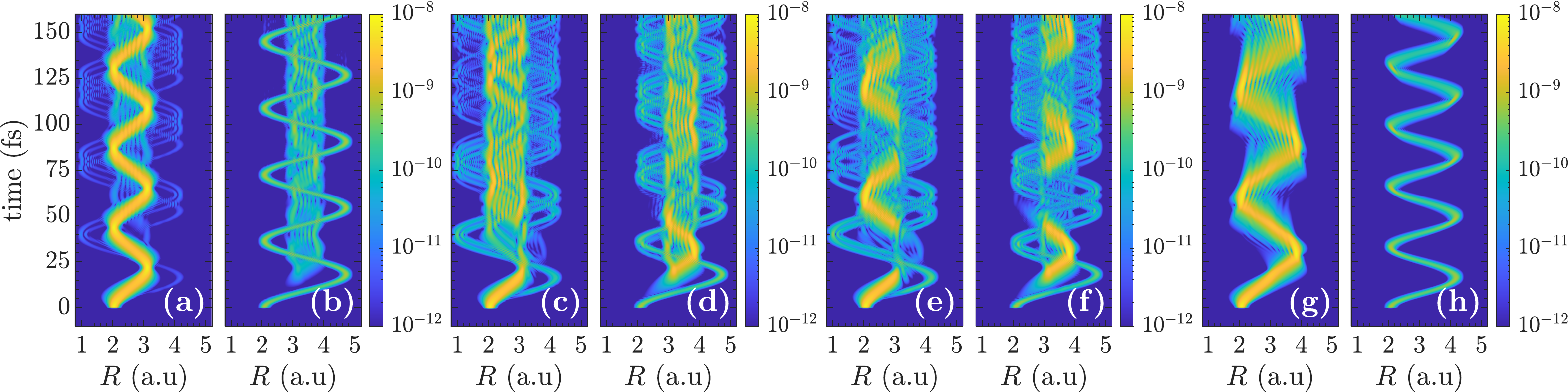}
		\caption{Nuclear dynamics on the two lowest excited curves. The figures show $\vert \chi_1(R,t) \vert^2$ and $\vert \chi_2(R,t) \vert^2$ in single-photon XUV-only spectroscopy with the incident XUV pulse arriving at time $t=0$. (a)--(b)~Dynamics on the diabatic curves with couplings in the weak regime (i.e., $\vert \chi_1^{(d)}(R,t) \vert^2$ and $\vert \chi_2^{(d)}(R,t) \vert^2$), $V_{12,0}^{(d)} = 0.0275$~eV, (c)--(d)~Dynamics on the diabatic curves with couplings in the intermediate regime, $V_{12,0}^{(d)} = 0.1649$~eV, (e)--(f)~Dynamics on the diabatic curves with couplings in the strong regime, $V_{12,0}^{(d)} = 0.2542$~eV, and (g)--(h)~Dynamics on the adiabatic curves without the nonadiabatic couplings in the strong regime (i.e., $\vert \chi_1^{(a)}(R,t) \vert^2$ and $\vert \chi_2^{(a)}(R,t) \vert^2$).}
		\label{Fig:NucDynam}
		\end{figure*}

For low amplitude of the diabatic coupling (e.g., $V_{12,0}^{(d)} = 0.0275$~eV), we are in the weak coupling regime, where the nuclear wave packets move almost exclusively on the diabatic curves, and the spectrum shows well-separated absorption lines. Comparing the spectrum with the vibrational levels from the diabatic electronic state $\ket{\Phi_1^{(d)}}$, we see that this agrees with a vertical Franck-Condon transition from the ground state to the first excited state $\ket{\Phi_1^{(d)}}$, see Fig.~\ref{Fig:PotentialsVibLevels}~(a). In the intermediate regime (e.g., $V_{12,0}^{(d)} = 0.1649$~eV), there is a significant coupling between the diabatic states, and the absorption spectra show a more irregular behavior. In this regime, neither the diabatic coupling nor the corresponding nonadiabtic coupling can be neglected, and transitions between the excited curves will occur in both representations. In this case the picture of the spectrum being a simple resolution of the vibrational levels of one of the excited state curves fails. In the strong coupling regime (e.g., $V_{12,0}^{(d)} = 0.2542$~eV), the diabatic curves are strongly coupled. Due to the strong coupling, the nuclear wave packet will make a Landau-Zener transition~\cite{bransden1992charge} between the diabatic curves as it passes through the crossing, which is equivalent to the motion being uncoupled on the corresponding adiabatic curves where the weak nonadiabatic coupling can be ignored. This part of the spectrum agrees with the the vibrational levels of the adiabatic state $\ket{\Phi_1^{(a)}}$, again stemming from a vertical Franck-Condon transition from the ground state, see Fig.~\ref{Fig:PotentialsVibLevels}~(b).

In Fig.~\ref{Fig:XUVonlyScan}~(b) we further show the XUV-only scan of the adiabatic curves without the nonadiabatic coupling. In this case there is no transition between the adiabatic states regardless of the strength of the diabatic coupling, $V_{12,0}^{(d)}$. The spectrum therefore reflects the vibrational levels of the first excited adiabatic state, $\ket{\Phi_1^{(a)}}$, populated by a Franck-Condon transition. Comparing the two scans in Fig.~\ref{Fig:XUVonlyScan} also highlights how the nonadiabatic coupling cannot be neglected in both the weak and intermediate diabatic coupling regime, but can be ignored in the strong diabatic coupling regime.

The dynamics on the two lowest excited electronic states $\ket{\Phi_1}$ and $\ket{\Phi_2}$ are shown in Fig.~\ref{Fig:NucDynam}, which depicts the norm squared of the two nuclear wave packets $\vert \chi_1 \vert^2$ and $\vert \chi_2 \vert^2$, see Eq.~\eqref{eq:NucExpansion}. Here we can explicitly see how the nuclear wave packets move almost uncoupled on the diabatic states in the weak regime in [Figs.~\ref{Fig:NucDynam}~(a) and \ref{Fig:NucDynam}~(b)], but couple in both the intermediate, [Figs.~\ref{Fig:NucDynam}~(c) and \ref{Fig:NucDynam}~(d)], and the strong, [Figs.~\ref{Fig:NucDynam}~(e) and \ref{Fig:NucDynam}~(f)], regimes. In the weak regime [Figs.~\ref{Fig:NucDynam}~(a) and \ref{Fig:NucDynam}~(b)] the main part of the excited wave packet is in the first excited electronic state, $\ket{\Phi_1^{(d)}}$, where the wave packet oscillates with a period of $\sim 40$~fs equivalent to the period of the vibrational levels $T = 2\pi/\omega_1 = 43$~fs of the first excited electronic state, see Eq.~\eqref{Eq:HarmExcited}. We further see how the adiabatic wave packets [Fig.~\ref{Fig:NucDynam}~(g) and \ref{Fig:NucDynam}~(h)] move in a manner, which is comparable to the dynamics of the diabatic wave packets in the strong regime if one takes into account the Landau-Zener transition~\cite{bransden1992charge}. For the adiabatic state [Figs.~ \ref{Fig:NucDynam}~(g) and \ref{Fig:NucDynam}~(h)] the wave packet is mainly on the first excited state, where the period of oscillation is $\sim 60$~fs.The adiabtic potentials are anharmonic, and the period is determined by using the average frequency $\omega_{\text{avg}}$ of the vibrational states seen in the spectrum in Fig.~\ref{Fig:XUVonlyScan}, giving $T = 2\pi/\omega_{\text{avg}} = 66$~fs. The increase in the period and the amplitude of the vibrations is a sign of the transition of the dynamics of the nuclear wave packets from primarily diabatic to adiabatic as the diabatic coupling strength is increased. Comparing the diabatic and adiabatic nuclear dynamics in the strong regime [Figs.~\ref{Fig:NucDynam}~(e)--(h)] one would assume that the sum over the two nuclear densities would be equal. There is, however, a small difference in the dynamics due to the ignored nonadiabatic coupling for the adiabatic curves, but the general motion agrees and the differences seen are one or two orders of magnitude lower than the main part of the wave packet.

\subsection{Full ATA spectra}
\label{Sec:FullATAS}

	\begin{figure}
		\centering
		\includegraphics[width=0.385\textwidth]{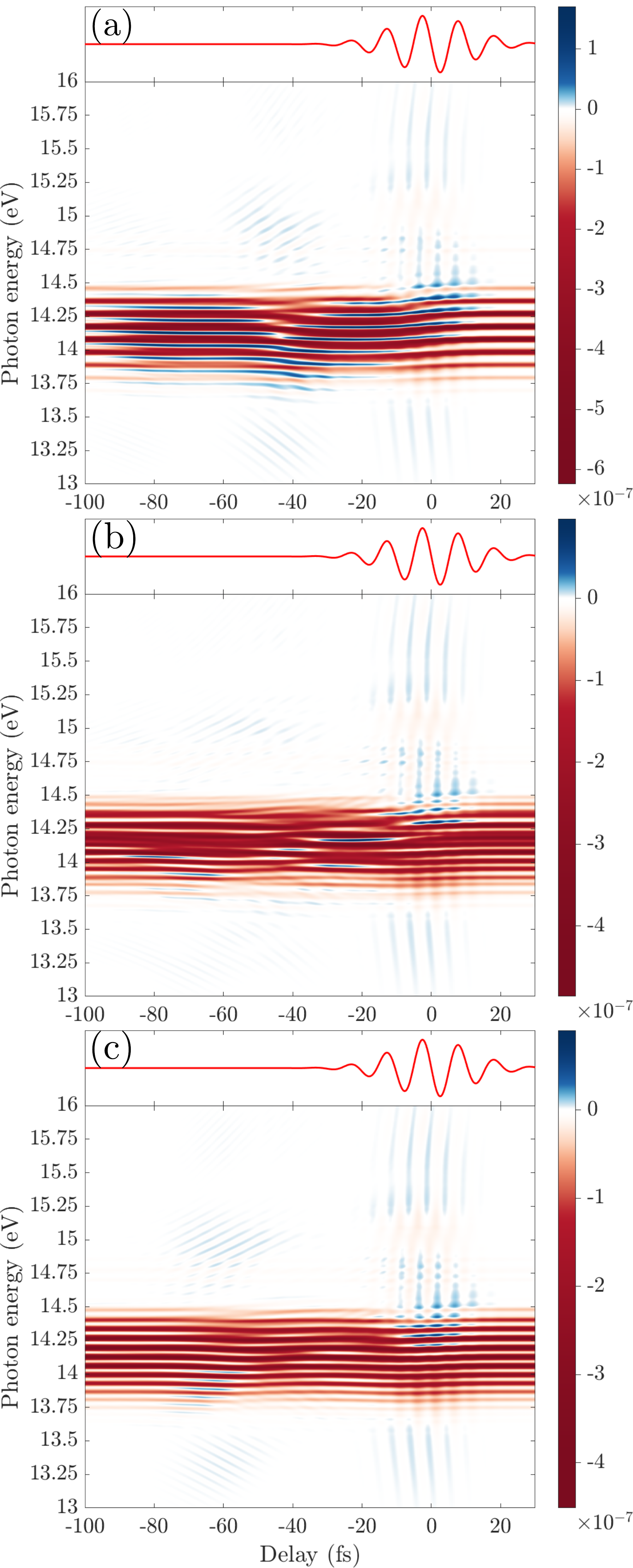}
		\caption{Full numerical ATA spectra calculated using Eq.~\eqref{eq:ResponseFunc} with the total dipole moment in Eq.~\eqref{eq:TimeDepDipole} obtained from a full numerical solution of dynamics using the exact diabatic curves. The top panel in each subfigure shows the IR pulse centered at $\tau = 0$~fs. (a) weak regime ($V_{12,0}^{(d)}=0.0275$~eV), (b) intermediate regime ($V_{12,0}^{(d)}=0.1649$~eV), (c) strong regime ($V_{12,0}^{(d)}=0.2542$~eV).}
		\label{Fig:ATAS_collected}
	\end{figure}

We now consider the full ATA spectra of the model system. To investigate the nonadiabatic effects we consider three distinct sets of curves corresponding to one of each of the three distinct regimes described in the analysis of the XUV-only spectra above [see Fig.~\ref{Fig:XUVonlyScan}]. 

In Fig.~\ref{Fig:ATAS_collected}~(a) we show the ATA spectrum following simulation of dynamics on the exact diabatic curves in the weak regime ($V_{12,0}^{(d)}=0.0275$~eV). For positive delays, the XUV pulse arrives after the IR pulse. If, for positive time delays, enough time passes, such that there is no temporal overlap between the two pulses, the resulting spectra are equal to the XUV-only spectra and we just see the delay-independent main absorption lines similar to Fig.~\ref{Fig:XUVonlyScan}. For large negative delays, the two pulse are again well separated in time. Here the XUV pulse arrives first and excites population onto the two bright state curves where the resulting wave packets will move, and later the incoming IR pulse will perturb these states giving rise to the delay-dependent features. For the largest negative delays, the window function, see Eq.~\eqref{eq:WindoWFunc}, will have damped most of the signal and we mainly see modulation of the main absorption lines. The emission features (blue) seen between the absorption lines could be due to hyperbolic sidebands~\cite{PhysRevA.96.013430}, but they are hard to identify clearly due to the closely-lying absorption lines (red) of the vibrational states. Features at zero delay, extending from the full width of energies considered, are attributed to the oscillating fringes, situated two IR photon energies away from the bright states $\sim 14 \text{ eV} \pm 2\times 0.4 \text{ eV} = (13.2, 14.8) \text{ eV}$, and the LIS, situated one IR photon energy away from the dark states at $\sim 15.5 \text{ eV} - 0.4 \text{ eV} = 15.1 \text{ eV}$. Both features extend over a range of energies, and as they both oscillate with twice the frequency of the IR, they blend together. In an atomic system, the oscillating fringes will extend continuously to negative delay, but in our case the nuclear dynamics moves the excited wave packets away from the Franck-Condon window of the ground state. This results in the fringes being present at zero delay and in a revival at $\sim -40$~fs. This time reflects the period of the nuclear oscillations seen for the lowest excited diabatic state $\ket{\Phi_1^{(d)}}$ in the weak regime in Fig.~\ref{Fig:NucDynam}~(a).

Figure~\ref{Fig:ATAS_collected}~(b) shows the spectra following simulation of the dynamics on the exact diabatic curves in the intermediate regime ($V_{12,0}^{(d)}=0.1649$~eV). The main absorption lines are more irregular compared with the result in the weak regime. This behavior is in accordance with what we saw in the XUV-only spectra in Fig.~\ref{Fig:XUVonlyScan}~(a). The closely-lying main lines also make it harder to distinguish any emission features between them. Similarly to the weak regime, we see features of the LIS and the oscillating fringes at zero delay. The fringes at negative delay are now situated at $\sim - 50$ fs instead of $\sim - 40$ fs, which can be attributed to a more significant mixing of the diabatic states leading to a larger period of the nuclear oscillations, see Figs.~\ref{Fig:NucDynam}~(c)-(d).

In Fig.~\ref{Fig:ATAS_collected} (c) we see the ATA spectra following simulation of dynamics on the exact diabatic curves in the strong regime ($V_{12,0}^{(d)}=0.2542$~eV). As expected from the XUV-only spectra in Fig.~\ref{Fig:XUVonlyScan}, the main absorption lines are more well separated compared to the intermediate regime, but we note weaker interlying emission features compared to the weak regime. We still see the LIS and oscillating fringes at zero delay. The fringes at negative delay have moved even further out at $\sim - 60$ fs, as the nuclear wave packets now move exclusively on the much broader adiabatic curve.

A calculation of the ATA spectrum following simulation on the adiabatic curves, without the nonadiabatic coupling, in the strong diabatic coupling regime produce results indistinguishable to the corresponding result for the diabatic curves in Fig.~\ref{Fig:ATAS_collected}~(c), as expected from the discussion of the results in Figs.~\ref{Fig:XUVonlyScan}--\ref{Fig:NucDynam}.

\section{Analytical multilevel model}
	\label{Sec:Analytical}

	\begin{figure}
		\centering
		\includegraphics[width=0.35\textwidth]{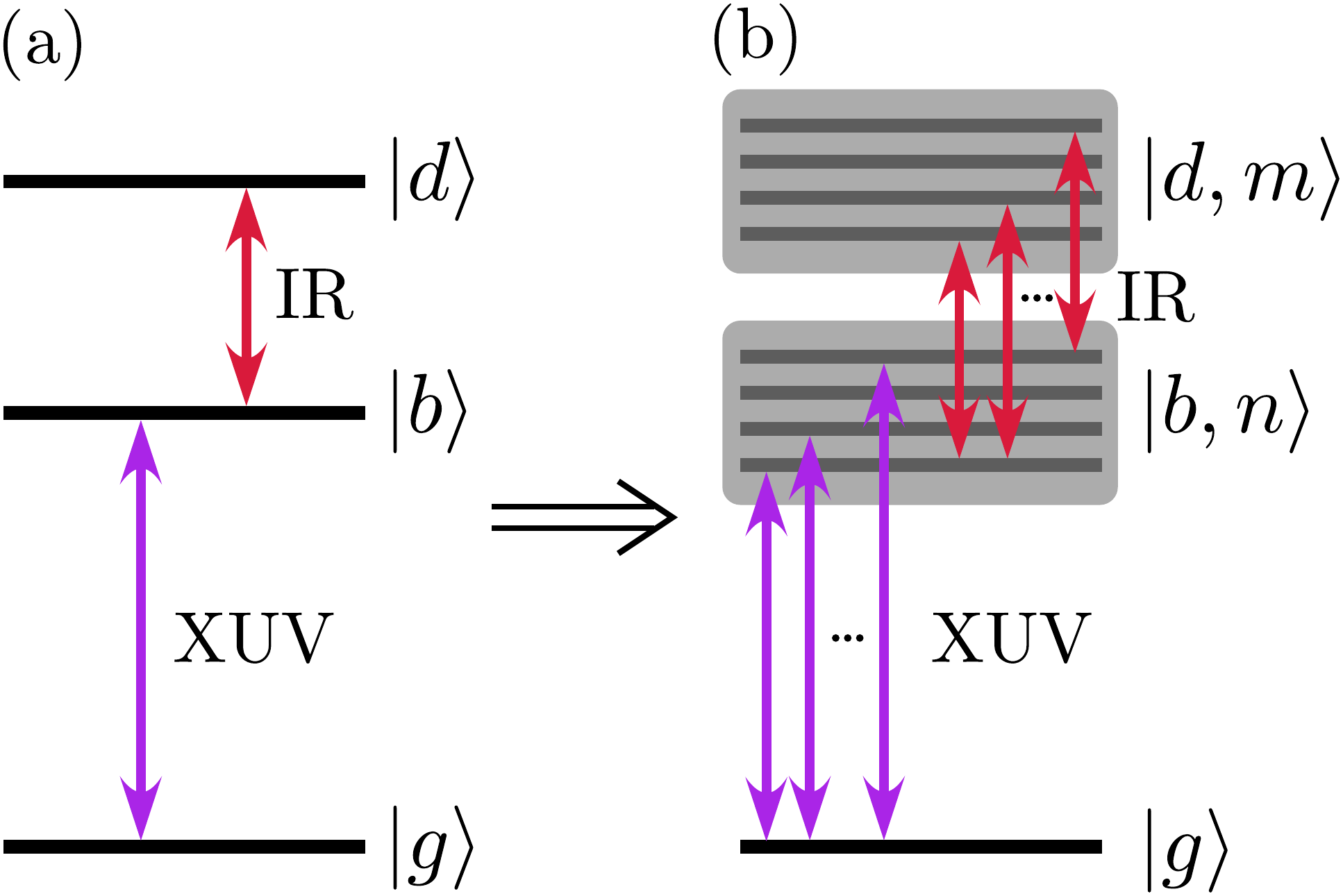}
		\caption{(a) Sketch of the level scheme used in the adiabatic three-level model. The model consists of a ground state $\ket{g}$ which couples to a bright state $\ket{b}$ through the XUV pulse, and a dark state $\ket{d}$, which is coupled to the bright state by the IR pulse. (b) Sketch of the level scheme used in the extended multilevel model capturing aspects of vibrational motion in the excited states by considering a manifold of bright $\ket{b,n}$ and dark $\ket{d,m}$ states.}
		\label{Fig:ThreeLevelModel}
	\end{figure}

To analyze ATA spectra in atoms, an adiabatic three-level model was used in Ref.~\cite{PhysRevA.96.013430} to derive an analytical expression for the response function in Eq.~\eqref{eq:ResponseFunc} describing the hyperbolic sidebands, oscillating fringes and LIS. In the three-level model one considers a ground state $\ket{g}$ and two excited states $\ket{b}$ and $\ket{d}$ which are bright and dark with respect to $\ket{g}$, respectively, with the coupling given by the XUV pulse. The two excited states are further coupled by the IR pulse, see Fig.~\ref{Fig:ThreeLevelModel}~(a). By approximating the short XUV pulse by an instantaneous delta function $F_{\text{XUV}}(t) \rightarrow \gamma\delta(t-\tau) $ and treating the interaction of the bright and dark states with the slowly-varying IR pulse adiabatically in time, it is possible, by working to second order in the field strength of the two pulses to obtain analytical expressions for the response function. 
A modification of the model to two levels with a permanent dipole was used in Ref.~\cite{PhysRevA.98.053401} to describe ATAS in the context of a polar diatomic molecule exemplified by the LiF system. The two-level model was further studied in Ref.~\cite{Drescher_2020} for an atomic system equipped with a permanent dipole.

To describe our model system we extend the three-level model, by considering a manifold of bright states $\ket{b,n}$ consisting of the vibrational states of the two lowest electronic states $\ket{\Phi_1^{(d)}}$ and $\ket{\Phi_2^{(d)}}$ together with a manifold of dark states $\ket{d,m}$ consisting of the vibrational states of the third excited electronic state $\ket{\Phi_3^{(d)}}$, see Fig.~\ref{Fig:ThreeLevelModel}~(b). The purpose of including more vibrational states is to capture effects of the vibrational motion. We then generalize the formulas for the spectrum based on the three-level model Ref.~\cite{PhysRevA.96.013430} by including the sum of the contributions of all sub three-level systems constructed from a sum over vibrational states of the three excited electronic states. The vibrational states are obtained by diagonalizing the field-free diabatic Hamiltonian in Eq.~\eqref{eq:ExactDiabHam}, and we denote their dipole moments by $D_{g;b,n} = \braket{g \vert \hat{D} \vert b,n}$ and $D_{b,n;d,m} = \braket{b,n \vert \hat{D} \vert d,m}$.

	\begin{table}
	\caption{Fitted values of the global line widths $\Gamma$ of the bright vibrational states and the delta function amplitudes $\gamma$ of the instantaneous XUV pulse in the three different regimes of diabatic coupling, see Eq.~\eqref{eq:ResponseMainLine}.}
	\label{Table:ResponseParams}
	\begin{ruledtabular}
	\begin{tabular}{lcc}
								&	$\Gamma$ [eV]	& 	$\gamma$ [a.u.]				\\ \colrule
		Weak regime, $V_{12,0}^{(d)} = 0.0275$~eV			&	0.0216			&	$3.93\times 10^{-4}$		\\
		Intermediate regime, $V_{12,0}^{(d)} = 0.1649$~eV	&	0.0222			&	$3.81\times 10^{-4}$		\\
		Strong regime, $V_{12,0}^{(d)} = 0.2542$~eV 			&	0.0214			&	$3.83\times 10^{-4}$
	\end{tabular}
	\end{ruledtabular}
	\end{table}

The response function describing the main absorption lines of the bright states
	\begin{equation}
		S_0(\omega,\tau)
%			= - \frac{2\gamma^2\rho}{c} \sum_n D_{g;b,n}^2 \omega \text{Im} \bigg[ \frac{1}{(\omega-E_{b,n})-i\Gamma/2} \bigg]
			= - \frac{2\gamma^2\rho}{c} \sum_n D_{g;b,n}^2 \omega \frac{\Gamma/2}{(\omega-E_{b,n})^2 + (\Gamma/2)^2}
			\label{eq:ResponseMainLine}
	\end{equation}
is obtained using only the instantaneous XUV pulse, see the Appendix. To avoid infinitely sharp lines we have further introduced a global finite line width $\Gamma$ for the bright state vibrational energies $E_{b,n} \rightarrow E_{b,n} - i\Gamma/2$. The values of the line width $\Gamma$ and the delta function amplitude $\gamma$ of the XUV is determined for each value of the diabatic coupling $V_{12,0}^{(d)}$ by fitting Eq.~\eqref{eq:ResponseMainLine} with the XUV-only spectrum of the full numerical model. The values are collected in Table~\ref{Table:ResponseParams}.
	
\begin{figure}
		\centering
		\includegraphics[width=0.45\textwidth]{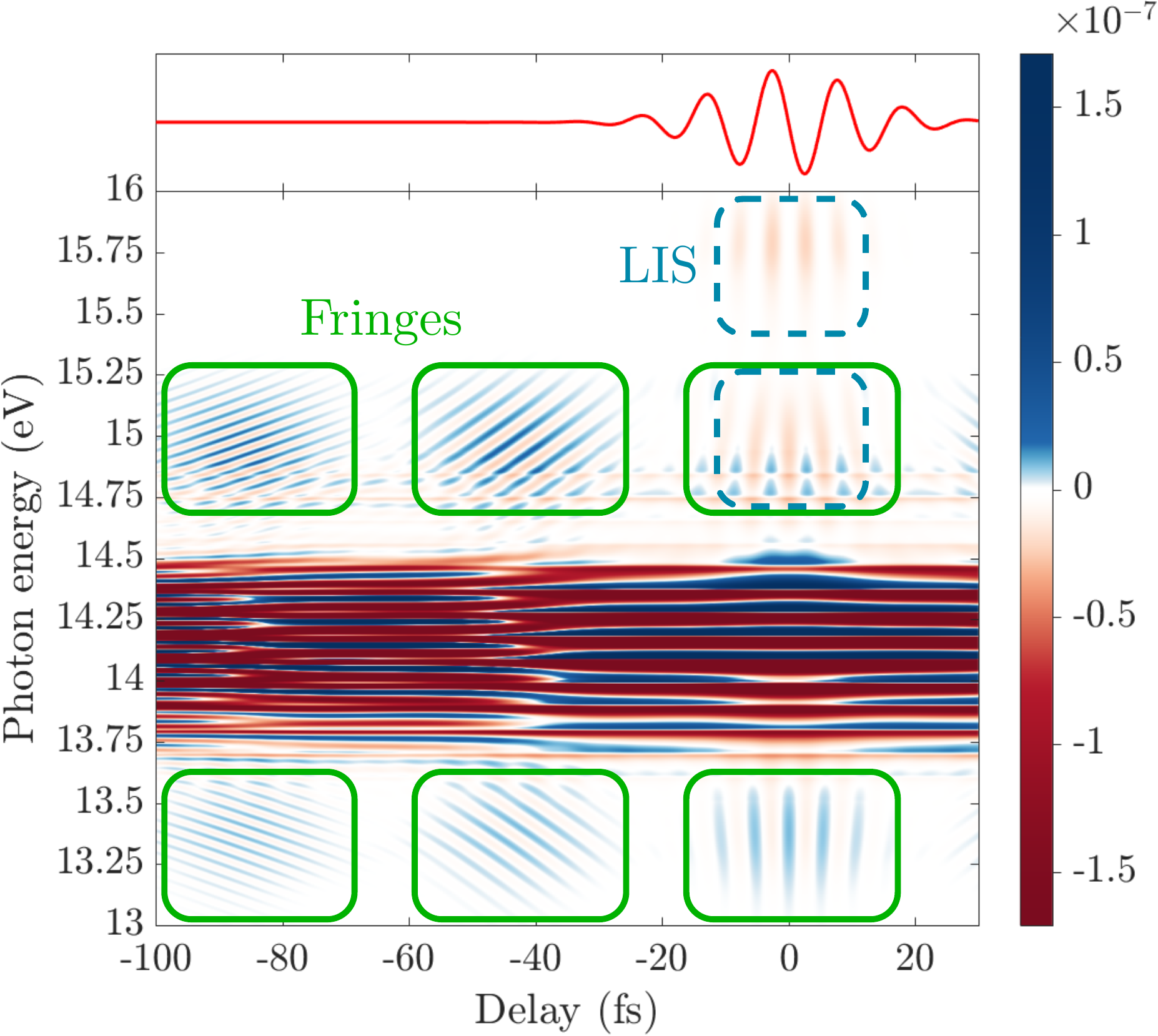}
		\caption{ATA spectrum for the diabatic curves in the weak regime ($V_{12,0}^{(d)}=0.0275$~eV) calculated using the multilevel model in Eqs.~\eqref{eq:ResponseMainLine}, \eqref{eq:ResponseOFandSB}, and \eqref{eq:ResponseLIS}. The positions of the oscillating fringes are indicated by the green solid boxes approximately $\pm 2\omega_{\text{IR}} = \pm 0.772$ eV around the main lines of the bright states, while the position of the LIS is indicated by the dashed blue boxes approximately $\pm\omega_{\text{IR}} = \pm 0.386$ eV around the dark states. The hyperbolic sidebands are situated between and on top of the main absorption lines. The top panel shows the IR pulse centered at $\tau = 0~\text{fs}$. To make the weaker features more visible, absorption in excess of the minimum value $|-1.6 \times 10^{-7}|$ is plotted in the same red color as that for the minimum.
}
		\label{Fig:ResponseFeatures}
	\end{figure}

\begin{figure*}
		\centering
		\includegraphics[width=0.8\textwidth]{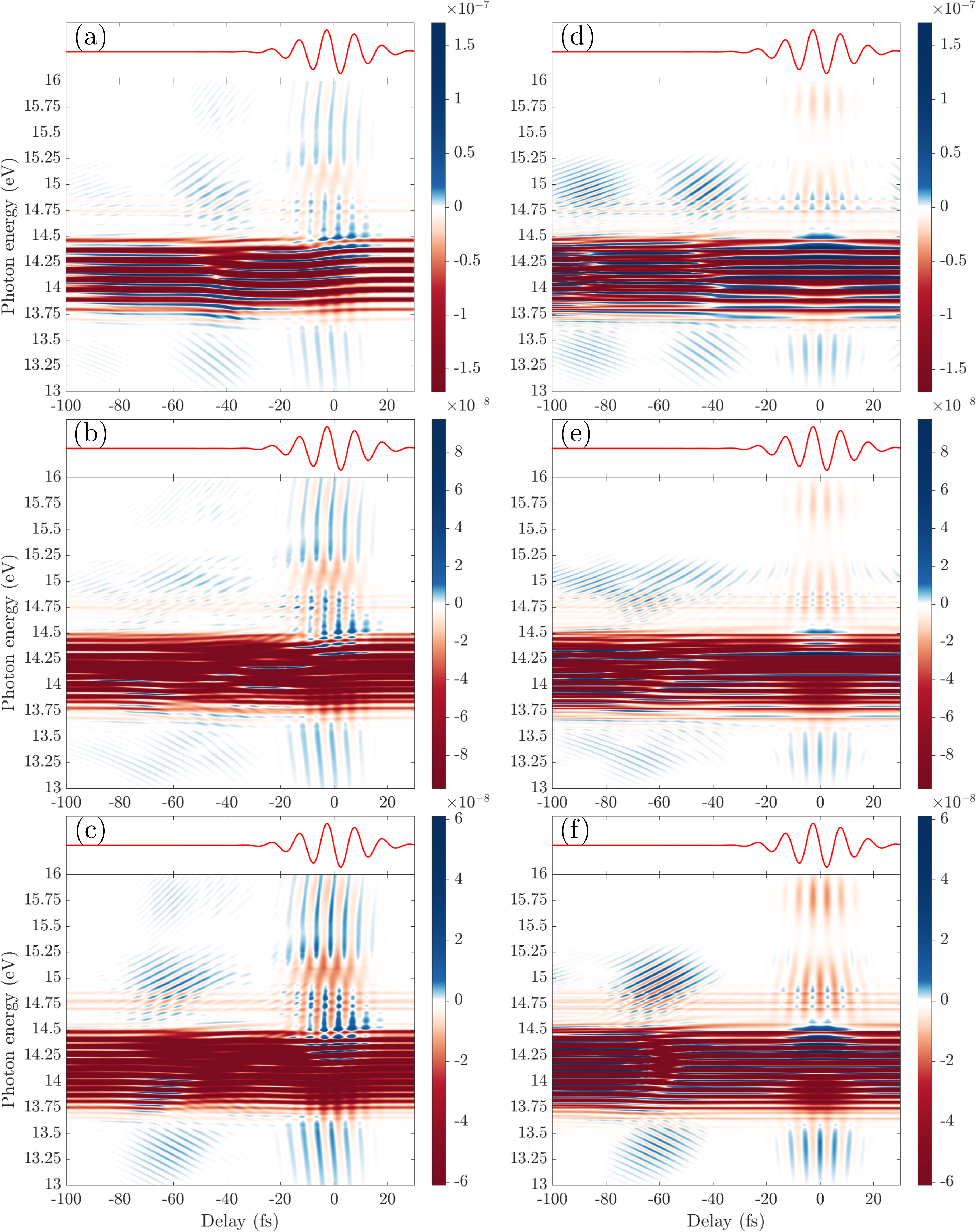}
		\caption{Comparison of ATA spectra of the full numerical model (a)--(c) with spectra obtained through the multilevel model of Eqs.~\eqref{eq:ResponseMainLine}, \eqref{eq:ResponseOFandSB}, and \eqref{eq:ResponseLIS} (d)--(f). The top panel of each subfigure shows the IR pulse centered at $\tau = 0~\text{fs}$. (a) and (d) weak regime ($V_{12,0}^{(d)}=0.0275$~eV), (b) and (e) intermediate regime ($V_{12,0}^{(d)}=0.1649$~eV), (c) and (f) strong regime ($V_{12,0}^{(d)}=0.2542$~eV). To make the weaker features more visible, absorption in excess of the minimum value $|-1.6 \times 10^{-7}|$ is plotted in the same red color as that for the minimum.}
		\label{Fig:AnalyticalResponse}
	\end{figure*}

Using the result of Ref.~\citep{PhysRevA.96.013430} and taking the sum over bright and dark vibrational states we find the following expression for the response function of the oscillating fringes and the hyperbolic sidebands
	\begin{widetext}
		\begin{align}
			S_1(\omega,\tau) &= -\frac{\sqrt{\pi}\gamma^2\rho F_{0,\text{IR}}^2 T_{\text{IR}}}{4\sqrt{2}c}
				\sum_m \sum_m \frac{D_{g;b,n}^2D_{b,n;d,m}^2}{E_{d,m}-E_{b,n}}
				\Bigg[ \frac{1}{\omega-E_{b,n}} + \frac{1}{2(E_{d,m}-E_{b,n})} \Bigg]
				\omega \cos \big[ (E_{b,n}-\omega)\tau \big]
			\label{eq:ResponseOFandSB} \\
				&\quad \times \Bigg[
				\exp\Bigg( -\frac{T_{\text{IR}}^2(\omega-E_{b,n}-2\omega_{\text{IR}})^2}{32} \Bigg)
				+\exp\Bigg( -\frac{T_{\text{IR}}^2(\omega-E_{b,n}+2\omega_{\text{IR}})^2}{32} \Bigg)
				- 2\exp\Bigg( -\frac{T_{\text{IR}}^2(\omega-E_{b,n})^2}{32} \Bigg) \Bigg]. \nonumber
		\end{align}
	\end{widetext}
The amplitude of the features is determined, among others, by the field strengths of the two pulses, $\gamma$ and $F_{0,\text{IR}} = \sqrt{I_{\text{IR}}}$, as well as the duration of the IR pulse, $T_{\text{IR}}$. The position in energy is determined by the three Gaussians. The term $(\omega - E_{b,n})^{-1}$ changes sign depending on whether $\omega < E_{b,n}$ or $E_{b,n} < \omega$ while the term $(E_{d,m} - E_{b,n})^{-1}$ is independent of $\omega$, leading to an asymmetry of the features around the bright states, with the features above the bright state energy being of highest amplitude. The first two Gaussians describe the oscillating fringes centered two IR energies away from the bright states, $\omega \approx E_{b,n} \pm 2\omega_{\text{IR}}$. At these positions in energy the cosine terms become $\cos \big[ (E_{b,n} - \omega) \tau \big] \simeq \cos [ 2\omega_{\text{IR}} \tau ]$, producing oscillations at twice the IR frequency as was seen in Fig.~\ref{Fig:ATAS_collected}. The oscillating fringes in Fig.~\ref{Fig:ATAS_collected} were further centered in time around zero delay, $\tau = 0$~fs, and with a revival at negative delay, $\tau \sim - 40$~fs to $\tau \sim - 60$~fs, respectively, indicating the period of the nuclear wave packet on the excited states. This revival time cannot be seen directly in Eq.~\eqref{eq:ResponseOFandSB}, but can emerge as a collective effect due to the interference of the terms as we will later see in Figs.~\ref{Fig:ResponseFeatures} and~\ref{Fig:AnalyticalResponse}. In the case of weak diabatic coupling, the bright states populated are approximately equidistant in energy, with the difference in energy $E_{b,n+1} - E_{b,n} \approx \omega_1$ given by the harmonic frequency of the electronic curve in Eq.~\eqref{Eq:HarmExcited}. The different cosine terms $\cos \big[ (E_{b,n} - \omega) \tau \big]$ will therefore be in phase and interfere constructively with a period $T = 2\pi/\omega_1 = 43$~fs, i.e. the exact revival time discussed earlier in Sec.~\ref{Sec:XUVscan}. In the case of strong diabatic coupling, the bright states considered are again approximately equidistant $E_{b,n+1} - E_{b,n} \approx \omega_{\text{avg}}$ separated by the anharmonic average frequency, giving constructive interference at $T = 2\pi/\omega_{\text{avg}} = 66$~fs. In the case of intermediate diabatic coupling, the bright states are more irregular and the revival is less pronounced with a period in between that of the weak and strong regime as was discussed in Sec.~\ref{Sec:FullATAS}. The last Gaussian describes the hyperbolic sidebands centered around the bright states $\omega \approx E_{b,n}$. At these energies the cosines $\cos [ \delta \tau ]$ oscillate with the small detuning $\delta = E_{b,n} - \omega \ll 1$, leading to almost no modulation at fixed photon energy, but instead give rise to hyperbolic shapes traced out by the curves $\delta \tau = \text{const}$. This hyperbolic behaviour is also present in the oscillating fringes, but to a much weaker degree. Naively the sidebands diverge at the position of the bright states due to the pole $(\omega - E_{b,n})^{-1}$. Again this can be eliminated by using the finite line width $\Gamma$ of the bright states. In principle, this should be introduced everywhere in Eq.~\eqref{eq:ResponseOFandSB}, but due to the small size of $\Gamma$ compared to the other energies included, it is only significant at the pole and we will only include it there for the sidebands. We further note, that since the energies $E_{b,n}$ enter in the derivation of Eq.~\eqref{eq:ResponseOFandSB} through a complex conjugate term, one should use the conjugate substitution $(E_{b,n})^* \rightarrow E_{b,n} + i\Gamma/2$ and then take the real part, i.e. $(\omega - E_{b,n})^{-1} \rightarrow (\omega - E_{b,n})/((\omega-E_{b,n})^2 + (\Gamma/2)^2)$.

Similarly we can obtain, by taking the sum over the result from Ref.~\cite{PhysRevA.96.013430}, the following expression for the response responsible for the LIS
	\begin{widetext}
		\begin{align}
			S_2(\omega,\tau) &= \frac{\sqrt{\pi}\gamma^2\rho F_{0,\text{IR}}^2 T_{\text{IR}}}{2c}
				\sum_n \sum_m \frac{D_{g;b,n}^2 D_{b,n;d,m}^2}{(E_{d,m}-E_{b,n})^2}
				\exp \Bigg( - \frac{4\tau^2}{T_{\text{IR}}^2} \Bigg)
				\omega \sin \big( \omega_{\text{IR}}\tau \big) \sin \big[ (E_{d,m} - \omega)\tau \big] \nonumber\\
				&\quad \times \Bigg[
				\exp\Bigg( -\frac{T_{\text{IR}}^2(\omega-E_{d,m}-\omega_{\text{IR}})^2}{16} \Bigg)
				-\exp\Bigg( -\frac{T_{\text{IR}}^2(\omega-E_{d,m}+\omega_{\text{IR}})^2}{16} \Bigg) \Bigg].
			\label{eq:ResponseLIS}
		\end{align}
	\end{widetext}
The LIS are situated in energy one IR frequency away from the dark states $\omega \approx E_{d,m} \pm \omega_{\text{IR}}$, and in time centered around zero delay, i.e., the LIS are only present when the IR and the XUV pulses have a temporal overlap. The sine term, $\sin \big[ (E_{d,m} - \omega)\tau \big]$, changes sign depending in whether $\omega < E_{d,m}$ or $E_{d,m} < \omega$, but the sign is balanced by the difference in sign of the two Gaussians, leading to the features being symmetrical around the dark states. At the energies of the LIS, $\omega \approx E_{d,m} \pm \omega_{\text{IR}}$, the oscillating terms become $\sin(\omega_{\text{IR}}\tau) \sin \big[ (E_{d,m} - \omega)\tau \big] \approx \pm \frac{1}{2}(1 - \cos(2\omega_{\text{IR}} \tau))$, showing that similar to the oscillating fringes the LIS oscillates at twice the IR frequency as seen in Fig.~\ref{Fig:ATAS_collected}. 

For the model to be valid, the transition energy $|E_{d,m} - E_{b,n}|$ must be larger than the IR frequency. In the present case we consider the 30 lowest vibrational bright states and lowest 30 vibrational dark states and disregard all sub three-level systems where the difference is below $2.5 \omega_{\text{IR}}$ in the sums in Eqs.~\eqref{eq:ResponseOFandSB}, and~\eqref{eq:ResponseLIS}, ensuring that we still include the systems contributing the most, i.e., the systems with the largest values of the product of the dipole moments squared $D_{g;b,n}^2 D_{b,n;d,m}^2$. The results has been checked for convergence. If the restriction on the considered states is not taken into account,  bright and dark states that are very close in energy will dominate due to the near divergence of the terms $(E_{d,m} - E_{b,n})^{-1}$ and $(E_{d,m} - E_{b,n})^{-2}$ in Eqs.~\eqref{eq:ResponseOFandSB}, and~\eqref{eq:ResponseLIS}, leading to unphysical signals even though the corresponding dipole moments, $D_{g;b,n}$ and $D_{b,n;d,m}$ of these states are very small. In the derivation of the model further assumptions are made, which lead to the signal being symmetric in time around zero delay and therefore to incorrect results for positive delay. The window function, see Eq.~\eqref{eq:WindoWFunc}, is also neglected and this leads to signals at larger negative delay that are stronger in the multilevel model compared to the results of the full numerical calculation.

Figure \ref{Fig:ResponseFeatures} shows the ATA spectrum for the weak diabatic curves ($V_{12,0}^{(d)}=0.0275$~eV) calculated using the adiabatic multilevel model using Eqs.~\eqref{eq:ResponseMainLine}, \eqref{eq:ResponseOFandSB} and~\eqref{eq:ResponseLIS}. In the photon energy range 13.75--14.5~eV we see the dominating main absorption lines of the bright states together with the hyperbolic sidebands. Centered around 15 and 15.75 eV in energy and zero delay, we see the LIS indicated by the dashed blue boxes. As discussed above the LIS oscillate with twice the IR frequence and are symmetric around the dark state. The latter, however, cannot be seen as the LIS at 15~eV are situated on top of the fringes.  The fringes are present around 13.25 and 15~eV at zero delay, and they do indeed show the revival at $-45$ and $-90$~fs respectively. We further note the hyperbolic shape of the fringes.

Figure~\ref{Fig:AnalyticalResponse} shows ATA spectra calculated in the three different regimes of diabatic coupling strength using the full numerical model in (a)--(c) compared with the spectra obtained by using the analytical multilevel model in (d)--(e). Qualitatively there is good agreement between the full numerical model and the multilevel model in all three regimes, with the main absorption lines, LIS and the oscillating fringes situated at the same energies. The multilevel model captures some modulation of the main lines. The LIS show the same oscillations in the two approaches, but the blue emission features in the full numerical model are absent in the spectra from the multilevel model, and as we noted above, the spectrum of the multilevel model is, as opposed to the full numerical model, symmetrical around zero time delay. For the oscillating fringes, the multilevel model reproduces the revival of the signals after one period of the nuclear oscillations seen in Fig.~\ref{Fig:NucDynam}, and the features exhibit the correct oscillations of twice the IR frequency together with the hyperbolic shape. In the case of the weak diabatic coupling regime in Fig.~\ref{Fig:AnalyticalResponse}~(d), we further see an additional revival after two periods. This revival is absent in the spectrum from the full numerical model in Fig.~\ref{Fig:AnalyticalResponse}~(a), but the discrepancy can be explained by the lack of a window function in the multilevel model. The oscillating fringes are best captured in the multilevel model, compared to the full numerical model, for the weak and strong diabatic regime, Figs.~\ref{Fig:AnalyticalResponse}~(d) and~(f), with the largest deviation in the intermediate regime, Fig.~\ref{Fig:AnalyticalResponse}~(e). In both the weak and strong diabatic coupling regime, the nuclear dynamics on the two bright state curves can be described as almost uncoupled if one considers the appropriate representation, the diabatic representation for the weak regime and the adiabatic representation for the strong regime. For the intermediate regime there is a significant coupling in both representations, and the the nuclear dynamics is not described by simple uncoupled oscillations. The latter leads to a less pronounced revival of the fringes, see Fig.~\ref{Fig:AnalyticalResponse}~(b), which is even more spread out for the multilevel model in Fig.~\ref{Fig:AnalyticalResponse}~(e). This suggests that the multilevel model performs best when the bright states can be considered uncoupled in one of the two representations, which is the case in either the weak or the strong coupling regime. Moreover the results suggest that less well-defined revival of the fringes in the spectra of the full numerical model can be a signature of intermediate diabatic coupling.

\section{Conclusion}
	\label{Sec:Conclusion}

In this work, we have considered the ATA spectra of a diatomic model system with varying strengths of the diabatic coupling. The model includes the diabatic coupling through a Gaussian ansatz making it applicable over a large range of geometries compared to vibronic models defined by expansions around minima. Based on a scan over the strength of the coupling we can divide the dynamics into different regimes with individual characteristic features in the spectra. Compared to the atomic case, the presence of nuclear dynamics in the diatomic molecule is in general reflected by additional absorption and emission lines associated with vibrational levels. If these features are regular and smoothly varying as a function of time delay between the XUV pump and the IR probe, it can be a signature of dynamics taking place with either weak or strong diabatic couplings in the diabatic representation. If, on the other hand,  the ATA spectra show irregular behavior as a function of time delay, it can be a signature of nonadiabatic dynamics and the importance of accounting for the nonadiabatic coupling terms between different Born-Oppenheimer states in the adiabatic representation. The numerical results were compared with the results of an analytical model, which was established based on extension of an atomic-based model. The molecular extension, involving account of different vibrational levels, captured part of the dynamics for weak and strong diabatic coupling regimes. At intermediate vibronic coupling strengths, where neither the crude diabatic or the approximate adiabatic approach without nonadiabatic couplings, are accurate, the dynamics is more rich and a full numerical approach is needed for the combined vibrationally coupled electronic and nuclear dynamics.

\section*{Acknowledgement}

This work was supported by the Danish Council for Independent Research (Grant No. 7014-00092B). The numerical results were obtained at the Centre for Scientific Computing Aarhus (CSCAA).

\appendix*
\section{Response for main absorption lines}
\label{App:MainLines}

In this Appendix, we describe the derivation of Eq.~\eqref{eq:ResponseMainLine} of the main text. In the description of the main absorption lines, we consider only the interaction with the XUV pulse. Without the presence of the IR pulse the dark states are uncoupled to the rest of the system, since we only consider the XUV to first order. As the system is initially in the ground state $\ket{g}$, we can write the full state as
	\begin{equation}
		\ket{\psi(t)} = a_g(t) \ket{g} + a_b(t) \ket{b} e^{-i(E_{b} - i\Gamma/2)t},
		\label{App:eq:Ansatz}
	\end{equation}
where we have included a finite line width in the energy of the bright state $\ket{b}$.	Inserting Eq.~\eqref{App:eq:Ansatz} into the time-dependent Schr{\"o}dinger equation with time-dependent Hamiltonian $H(t) = H_0 - F_{\text{XUV}}(t) \hat{D}$ and projecting onto the two states $\bra{g}$ and $\bra{b}$, we obtain approximately the coupled equations for the expansion coefficients
	\begin{align}
		i\dot{a}_g (t) &= - F_{\text{XUV}}(t) D_{g;b} a_b(t) e^{-i(E_b -i\Gamma/2) t} \\ 
		i\dot{a}_b (t) &= - F_{\text{XUV}}(t) D_{g;b} a_g(t) e^{ i(E_b -i\Gamma/2) t}.
	\end{align}
Using an instantaneous delta function as an approximation for the XUV pulse $F_{\text{XUV}}(t) = \gamma \delta(t - \tau)$, and working to first order in the amplitude $\gamma$ with initial conditions $a_g = 1$ and $a_b = 0$ the equations above are readily integrated
	\begin{align}
		a_g (t) &= 1 \\
		a_b (t) &= i\gamma D_{g;b} \theta (t - \tau) e^{i (E_b - i\Gamma/2) \tau}.
	\end{align}
The time-dependent dipole moment of the state $\ket{\psi(t)}$ is
	\begin{align}
		\braket{D} \! (t) &= i \gamma D_{g;b}^2 \theta (t-\tau) e^{-i (E_b -i\Gamma/2) (t-\tau)} \\
			& \quad - i \gamma D_{g;b}^2 \theta (t-\tau) e^{ i (E_b +i\Gamma/2)(t-\tau)}.
	\end{align}
We neglect the window function, see Eq.~\eqref{eq:WindoWFunc}, and consider only positive frequencies as these are what enters into the spectrum
	\begin{equation}
		\braket{D} \! (t) = - i \gamma D_{g;b}^2 \theta(t-\tau) e^{i E_b (t-\tau)} e^{-\Gamma t/2}.
	\end{equation}
The Fourier transform
	\begin{equation}
		\tilde{D}(\omega,\tau)
			= -i\gamma D_{g;b}^2 \mathcal{F} \big[ \theta (t-\tau) e^{i E_b (t-\tau)} e^{-\Gamma (t-\tau)/2} \big](\omega)
	\end{equation}
can be performed by using the following two translation-phase shift properties of the Fourier transform
	\begin{align}
		\mathcal{F} [f(t-\tau)](\omega) 		&= \mathcal{F} [f(t)](\omega) e^{-i \omega \tau} \\
		\mathcal{F}[g(t)e^{i E_b t}](\omega) 	&= \mathcal{F}[g(t)](\omega - E_b),
	\end{align}		
together with the Fourier transform of the function
	\begin{equation}
		\mathcal{F}[\theta(t) e^{-\Gamma t/2}](\omega) = \frac{1}{\sqrt{2\pi}} \frac{1}{ i\omega + \Gamma/2 }
	\end{equation}
yielding
	\begin{equation}
		\tilde{D} (\omega,\tau)
			= -i \frac{\gamma D_{g;b}^2}{\sqrt{2\pi}} \frac{1}{i(\omega - E_b) + \Gamma/2} e^{-i\omega\tau}.
	\end{equation}
Inserting this into the expression for the response function in Eq.~\eqref{eq:ResponseFunc}, together with the Fourier transform of the XUV delta function $\tilde{F}_{\text{XUV}}(\omega) = 1/\sqrt{2\pi} e^{-i\omega\tau}$ we find
	\begin{equation}
		S_0(\omega,\tau)
			= - \frac{2\gamma^2\rho}{c} D_{g;b}^2 \omega \frac{\Gamma/2}{(\omega-E_{b})^2 + (\Gamma/2)^2}.
	\end{equation}
The response function in Eq.~\eqref{eq:ResponseMainLine} is then obtain by taking the sum over the different bright states.

%\bibliography{ArticleBib.bib}


\end{document}